\documentclass[conference]{IEEEtran}

\usepackage{cite}
\usepackage{amsmath,amssymb,amsfonts}
\usepackage{algorithmic}
\usepackage{graphicx}
\usepackage{textcomp}
\usepackage{verbatim}
\usepackage{color}
\def\BibTeX{{\rm B\kern-.05em{\sc i\kern-.025em b}\kern-.08em
    T\kern-.1667em\lower.7ex\hbox{E}\kern-.125emX}}
\begin{document}

\title{Quantifying Information Exposure\\ in Internet Routing}

\author{
\IEEEauthorblockN{Peter Mell}
\IEEEauthorblockA{\textit{National Institute} \\
\textit{of Standards and Technology}\\
Gaithersburg MD, USA \\
peter.mell@nist.gov}
\and
\IEEEauthorblockN{Assane Gueye}
\IEEEauthorblockA{\textit{Electrical Engineering} \\
\textit{University of Maryland, College Park}\\
\textit{University Alioune Diop of Bambey},\\
assane1.gueye@uadb.edu.sn
\and
\IEEEauthorblockN{Christopher Schanzle}
\IEEEauthorblockA{\textit{National Institute} \\
\textit{of Standards and Technology}\\
Gaithersburg MD, USA \\
christopher.schanzle@nist.gov}
}}

\maketitle

\begin{abstract}

Data sent over the Internet can be monitored and manipulated by intermediate entities in the data path from the source to the destination. For unencrypted communications (and some encrypted communications with known weaknesses), eavesdropping and man-in-the-middle attacks are possible. For encrypted communication, the identification of the communicating endpoints is still revealed. In addition, encrypted communications may be stored until such time as newly discovered weaknesses in the encryption algorithm or advances in computer hardware render them readable by attackers.

In this work, we use public data to evaluate both advertised and observed routes through the Internet and measure the extent to which communications between pairs of countries are exposed to other countries. We use both physical router geolocation as well as the country of registration of the companies owning each router. We find a high level of information exposure; even physically adjacent countries use routes that involve many other countries. We also found that countries that are well `connected'  tend to be more exposed. Our analysis indicates that there exists a \emph{tradeoff} between robustness and information exposure in the current Internet.
\end{abstract}


\section{Introduction}

Data sent over the Internet can be monitored and manipulated by intermediate entities in the path from the source to the destination. For unencrypted communications (and some encrypted communications with known weaknesses), eavesdropping and man-in-the-middle attacks are possible. For encrypted communication, the identification of the communicating endpoints is still revealed. This is important for certain security sensitive communications (e.g., communication between military commands and units). In addition, encrypted communications may be stored until such time as newly discovered weaknesses in the encryption algorithm or advances in computer hardware render them readable by attackers. This kind of attack is especially dangerous as quantum computers, that can break widely used public key encryption, become a reality.

This work is an attempt to quantify this \emph{global information exposure} in the Internet by measuring the extent to which communications between pairs of countries are exposed to other countries. We focus on the routers relaying the packets in Internet communications and use two publicly available datasets to evaluate both advertised and observed routes through the Internet.

With the first dataset, we focus on the physical geolocation of the routers. Every router resides within a unique national boundary and is required to operate according to the laws of that nation. Thus, the data traversing the nation may be exposed to government eavesdropping or control. This dataset was essentially traceroute data from a worldwide collection of monitors (probing sites). We determined the country of residence of each router by geolocating it Internet Protocol (IP) address and used that to convert the router paths to country paths. This gives us a security model in which each router is mapped to their country of residence. Whether or not countries use due process and/or provide transparency for such data access does not affect our results. 

With the second dataset, we focus on the legal ownership of the routers. Every router is part of an autonomous system (AS) and every AS is owned by a company that has a country of registration. In this approach, we map each router to the country in which their AS is registered. This models companies abiding by the laws of their country of registration and providing access to the routers that they own. This may be required in some countries or optional in others for assets outside of the physical boundary of the country. These distinctions are irrelevant for our research as we are evaluating worst case data exposure. This dataset was from BGP router tables where we converted advertised routes to country paths through mapping ASs to their country of registration.

Using these two datasets we performed several experiments. In the first experiment we evaluated how well the data from a set of monitors (BGP routing tables or traceroute probing sites) in a country `generalized' to other sites in the country. This experiment undergirds the utility of the other measurements. We then measure the number of countries `involved' in communication between pairs of countries with respect to the distance between the paired countries. We next randomly choose increasing sets of untrusted countries to be `excluded' from communication exchanges. We saw how well pairs of countries could avoid that their communications transit through the excluded countries. Lastly, we perform graph centrality analyses (closeness, degree, eigenvalue, and load) on the graphs generated from using the country paths from both datasets.

Our generalization results showed that it is possible to use a small number of monitoring sites within a country, $x$, in order to represent the country to country network traffic of the rest of the sites in $x$ to a high degree of coverage.

With the `involved' country experiments, the geolocation and registration data produced very similiar results. We discovered that adjacent countries (on our country communication graph of the Internet) still have a high number of involved countries between them (those that can view their network traffic). Even more surprising, the number of involved countries actually increases as the country graph distance decreases. In general the number of involved countries was slightly higher for the registration data compared to the geolocation data. Lastly, we found that countries were extremely close together (usually just 1 or 2 hops for most countries). 

With the `excluded' country experiments, the results from the geolocation and registration data differed significantly. We show that it is insufficient to use just the geolocation data to determine information exposure, as has been done in previous work. We show that for a small number of excluded countries (e.g., less than 10), in general there is a high chance that a country can send data to another country where all known routes avoid the excluded countries. However, this likelihood decreases extremely rapidly with more than 10 excluded countries.

With the country graph centrality experiments, we compute the average number of involved countries (between a given country to all the other possible destinations) with respect to the centrality (closeness, degree, eigenvalue, and load) of that country. We show that countries with high centrality values (i.e., well connected) tend to have higher information exposure. This has been observed with all centrality metrics and with both datasets. This observation is consistent with the findings of the `involved' country experiments, where we discovered that adjacent countries  still have a high number of involved countries between them. Indeed, when a country has high (say degree) centrality, it has many direct neighbors. Since the number of involved countries is high for each neighbor, the average is consequently high. This seems to suggest that in the current Internet, there is a tradeoff between the ``connectivity'' of a country and its degree of information exposure. Indeed, a country that is well connected has many alternate paths to each destination (which is desirable for the robustness of routing). However, the diversity of paths also implies that many countries (some potentially adversarial)  might be traversed by the communications to a given destination. We are not aware of any other study revealing this robustness-exposure tradeoff.

\section{Data Description}
\label{data}

We obtained our data from the public datasets provided by the Center for Applied Internet Data Analysis (CAIDA) \cite{CAIDAData2017}, covering the years 2015 and 2016. We collected both Border Gateway Protocol (BGP) routing tables to view advertised AS routes through the Internet as well as traceroute type data from a worldwide set of monitors. We converted both AS and router paths into country paths (as described below). A challenge is that our datasources reveal Internet paths that are both advertised (registration data) and used (geolocation data), however, neither dataset reveals how often these paths are used. In addition, there likely exist additional routes not revealed from our data sources. Our experimental results thus are a lower bound on the extent to which information exposure is taking place at the country level. This said, it is reasonable to assume that our discovered routes cover the primary pathways through the Internet.

\subsection{Geolocated Router Path Data}

Our first dataset, which we call the `geolocated' data, consisted of actual paths discovered through active scamper probing \cite{Scamper2017} (similiar to traceroute) by a worldwide set of CAIDA Archipelago (Ark) monitors \cite{Ark2017}.
We collected all daily traces from January 01, 2015 to December 31, 2016 (a total of 123\,121 files totaling 2.3 TB). After pre-processing and duplicate removal, we ended up with more than 3.1 billion distinct probe traces for each year. We then used the MaxMind\footnote{Any mention of commercial entities or products is for information only; it does not imply recommendation or endorsement by NIST.} service \cite{Maxmind2017} to geolocate each router within a particular country and we converted the router paths into country paths. While geolocation data of routers can be inaccurate, previous work has found that it is more accurate at a country level of abstraction \cite{Huffaker2011}.

\subsection{Autonomous System Path Data}

Our second dataset, which we call the `registration' data, consisted of Border Gateway Protocol (BGP) routing tables from a worldwide set of routers. This provided advertised routes between autonomous systems (ASs). We obtained the data using the BGPStream tools \cite{BGPStream2017} to collect data from the University of Oregon Routeviews Project \cite{Routeviews2017} from Jan 01, 2015 to Dec 31, 2016 (a total of 150 GB of raw data). After pre-processing and duplicate removal, we ended up with more than 2.5 billion path traces for each year. Using other CAIDA provided data, we mapped ASs to their countries of registration thereby converting the AS paths to country paths.

\section{Data Generalization Experiment}
\label{generalization}
Both datasets yield country paths through the Internet originating from specific locations or `monitors' (a router that provided its BGP tables or a scamper probing site). Any particular country will have zero or more monitors. The monitors tend to be distributed throughout a country as there is little motivation to monitor the same location multiple times. Thus, we assume that the monitors have somewhat of a random distribution but acknowledge that this is not strictly true.

To undergird the results of our information exposure experiments, it is necessary to show that the set of monitors within particular countries provide sufficient data to represent the entire country (that they `generalize'). More specifically, the set of country paths yielded by the monitors should closely approximate the set of country paths that would be revealed in the hypothetical case of having monitors in every location within the country.

We approximately test this by comparing the paths revealed by each monitor with the paths revealed by all other monitors within a particular country. Let $M$ represent the set of monitors in a country. Let $R(W)$ represent the set of country paths revealed by the monitors in set $W$. Let $x \in M$ and $Y = M \setminus x$. For each $x$, we compute a ratio $|R(\{x\}) \cap R(Y)| / |R(\{x\})|$ which we refer to as the `generalization ratio'. We then plot the mean of all such computed ratios for each country against the number of monitors in that country. 

For the registration data, the mean generalization ratio increases very quickly to at least .7 with 20 monitors and around .9 with 60 monitors. One country had 3303 monitors and another 13\,912 monitors 
resulting in generalization ratios of .99. 
For the geolocation data, the mean generalization ratio also increases very quickly to at around .8 with just 5 monitors. One country had 44 monitors with a generalization ratio of around .96. These results show that with a sufficient number of monitors within a country, our data generalizes to represent the country paths used by a vast majority of the countries.

\section{Information Exposure Experiments and Analysis}
\label{experiments}

All experiments were performed on both datasets (geolocation and registration) for both 2015 and 2016. Due to space constraints, we limit ourselves to summarizing our findings from the data. Some example results are provided, anonymizing the countries as Bespin and Hoth. 

\subsection{Number of Countries Involved in Pairwise Communication}
\label{ss.involved}

Our first experiment is to measure how many countries are involved in country to country Internet communications. The set of `involved' countries, $I(x,y)$, for a pair of communicating countries $x$ and $y$ consists of all countries on any recorded country path from $x$ to $y$ (excluding both $x$ and $y$). The involved countries represent the minimal set of countries that could observe or modify some fraction of the communications from $x$ to $y$. Note that even with encrypted traffic, this measurement matters for certain high sensitivity communications (e.g., military commands) as the communicating endpoints are revealed.

\subsubsection{Experiment}

\begin{figure}
\centering
\includegraphics[scale=.45]{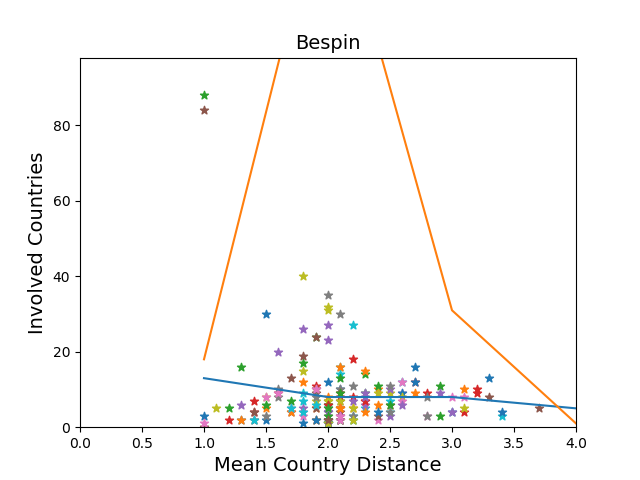}
\caption{Bespin Mean Involved Country Exposure from Geolocation data for 2016}
\label{fig.involved.2016-Mean-US-Geo}
\end{figure}

For our experiment, we calculated $I(x,y)$ for all paths between all pairs of countries. We did this for both the registration and geolocation data.
We evaluated each source country $x$ individually, creating a figure to analyze each country (an example is shown in figure \ref{fig.involved.2016-Mean-US-Geo}). For each $x$ we plotted each target country $y$ using a figure with an x-axis representing the mean country distance among all paths and the y-axis representing the total number of involved countries using all paths. We then performed the same analysis but used the minimum country distance for the x-axis.

\subsubsection{Discussion}

We find that the number of involved countries can be high. This is true for both datasets. One large wealthy nation has up to 96 involved countries between it and another country. Surprisingly, we find this even for countries that are adjacent. This indicates that even when countries have close proximity (either geographically or logically), the routing structure of the Internet will use many non-direct paths. While likely necessary and appropriate for dynamic load balancing, it has a huge effect in increasing the worst case information exposure between countries on the Internet.

Furthermore, we find a relationship between the number of involved countries and the distance between the countries (i.e., number of intervening countries). The number of involved countries generally decreases as the distance increases. This result seems counter intuitive and has a great impact on the evaluation of the privacy of communications between pairs of countries.

\subsection{Excluding Countries from Communications}
\label{ss.excluded}

In this experiment we evaluate how easily particular countries can communicate to all other countries without their communications traversing some target set of countries. 

\subsubsection{Experiment}
\begin{figure}
\centering
\includegraphics[scale=.45]{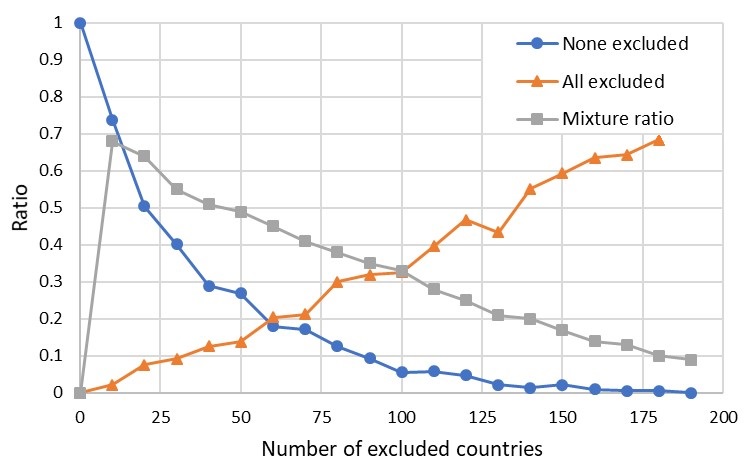}
\caption{Hoth Excluded Country Exposure from the Geolocation Data for 2016}
\label{fig.2016-BR-Geo}
\end{figure}

We execute 20k trials per country. We test excluded country list sizes ranging from 0 to 190 using a step of 10. For each size, we ran 500 trials; for each trial we choose a random destination country and a random set of countries to be on the excluded list for that trial. For each country we created a figure showing the mean ratio of paths containing no excluded countries over all paths (including data to all the other countries in a single figure). We also plotted the ratio for all paths containing excluded countries as well as when the paths contained a mixture (some having excluded countries and some not). An example figure is shown in figure \ref{fig.2016-BR-Geo}.

\subsubsection{Discussion}
For all countries, as the number of countries on the excluded list increases, the ratio of paths containing no excluded countries decreases rapidly (a roughly geometric or exponential decay depending upon the country). For one wealthy large country, an excluded list of size 10 using the geolocation data yields a 57\% chance of no paths having an excluded country (choosing some other country at random as the destination). However, increasing the list to 20 reduces the chance to just around 37\%; at 50 it is 10\%. These numbers vary dramatically given specific scenarios with set country pairs and should not be used to evaluate the overall state of privacy for Internet communications. 

Performing the same example evaluation using the registration data changes the results to around 9\%, 6\%, and 2\% respectively. This shows that evaluating information exposure using just the geolocation data is insufficient as the registration data has significantly different result values (although the similarly shaped functions). To our knowledge, all past research in this area has focused solely on using the geolocation data (or they used the router registration BGP data as a substitute for router geolocation, which we show in section \ref{related} to be flawed).

\subsection{Comparison to Country-Country Communication Graph}
\label{ss.comparisonmetrics}
We also analyzed the overall communication graphs using centrality metrics. To our knowledge, these are novel results as we focus on centrality measurements for the Internet as a whole from the perspective of country to country communications.

\subsubsection{Experiment}
In this experiment we take the following types of centrality metrics: closeness, degree, eigenvalue, and load. Degree centrality indicates the number of nodes connected to a given node. Closeness centrality measures the mean distance from a vertex to other vertices. Eigenvalue centrality is a measure of the structural importance of nodes, proportional to the structural importances of their connected neighborhood. The load centrality of a node is the fraction of all shortest paths that pass through that node.

\begin{figure}
\centering
\includegraphics[scale=.5]{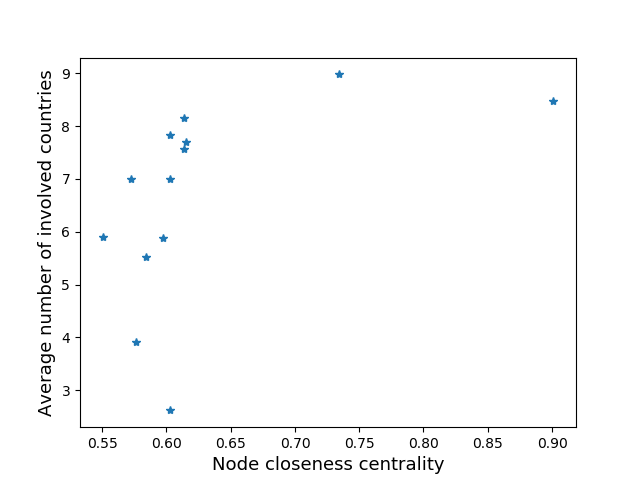}
\caption{Node Closeness Centrality for 2016 Geolocation Data}
\label{fig.graphc.geoloc.2016-cl}
\end{figure}

The experiment works as follows. First, we generate a country-level communication graph by: (1) merging all routers (or ASs) within a same country into one node and removing all loops; (2) considering links between routers (or ASs) in different countries as links between the nodes representing the countries (all multi-edges are removed).  Second, for each country (which represents a node in the graph) and for each centrality metric, we compute the average number of involved countries to all possible destinations from that country.  We finally scatter-plot the average number of involved countries as a function of the value of the centrality metrics. An example result is provided in figure \ref{fig.graphc.geoloc.2016-cl} (the rest are omitted due to space limitations. 

\subsubsection{Discussion}
\label{discussion}

We find that the general trend is that as the centrality values of the nodes increase, the average number of involved countries increases. In other terms, as a country is more connected, its information exposure increases in the sense that there are more countries that might be involved in a communication between that country and a random destination in the world.

The observations above are consistent with what we have found with our `involved' countries experiment. 
For recall, we observed that adjacent countries still have a high number of involved countries between them.
As involved countries are determined using routes discovered in the dataset, this is equivalent to saying that 
even for pairs of countries between which there is a direct communication path, there exist many alternate and independent paths that 
go through at least one other country. Hence, if a country has many neighbors (in the communication graph), 
there will exist many such alternate paths (going through third countries), and the average number of involved countries
will be large. This could also be explained by the dis-assortativity \cite{newman2002assortative} property of the Internet, which states that ``nodes with high centrality metric tend to be connected together''. A consequence of this is that those nodes will tend to form together a very well-connected sub-graph with many independent paths. 
On the other hand, as the distance between the two countries increases, there will be less of those independent paths, resulting to a lower average of involved countries.

Having many independent paths is by itself a well-desired property for the robustness of the Internet. 
Our study shows that there is an unintended cost to this robustness: \emph{the exposure of information}. Since the Internet routing is based on ``best effort'', any of the alternate paths can be used for a given communication between two countries. However, with more alternate paths between two countries, there is a greater risk that their communications transit via an un-trusted country, hence increasing their information exposure.
To our knowledge, this paper is the first study that reports on this tradeoff between robustness and information exposure.

\section{Related Work}
\label{related}

\cite{Edmundson2016} is most similar to our work in that they evaluate routes to determine their information exposure at a country level of abstraction. It uses traceroute and due to limitations of their measurement infrastructure, they limit their analysis to five countries. They focus on measurements of regular users accessing the Alexa Top 100 websites and provide related analytics on specific countries. In contrast, our work is focused on information exposure measurements for the movement of high sensitivity data of interest to foreign nation-states. Instead of being limited to just a few countries, our approach using publicly available datasets can be applied to all countries. Of perhaps greater importance is that our work did not just evaluate router geolocation dataset as in \cite{Edmundson2016}, it includes router country of registration data as well. 

\cite{Karlin2009} and \cite{Shah2015} use BGP data for an analysis of nation-state routing. However, they use BGP data to approximate the country of residence of each router on a path. At the AS level of granularity, the country location of the specific routers used in a path cannot be accurately determined \cite{Edmundson2016}. 
To easily see this, consider the Internet backbone provider Level 3. They own routers throughout the world, on every continent,
and yet their country of registration is the United States \cite{CAIDAData2017}. Using the Level 3 Internet backbone, communications paths can physically traverse the entire globe and yet appear using the BGP data to stay within the United States. We use the more accurate geolocated traceroute data for this purpose and uniquely use the BGP data to show the countries that have influence over the companies owning the routers (regardless of their physical location). Both perspectives are important and our work is the first to take both into account.

\cite{Mell2015} and \cite{Mell2015a} evaluate how groups of countries could collude to partition up the Internet into isolated chunks in order to prevent pairs of countries from communicating. The two papers use geolocated router paths converted to country paths as we do but focus on using them to measure the possibility of active attacks as opposed to measuring information exposure.

\section{Conclusion}
\label{conclusion}

We have quantified the information exposure in the global Internet with respect to countries having access to (or even modifying) data in transit between other countries. Our experiments covered all countries and we presented here the results for two representative countries. We have found that the level of exposure is significant. Even for communicating countries that are physically adjacent, many paths involving other countries are used. Physical proximity does not guarantee private communications. Our study has also shown that there is an apparent tradeoff between robustness and information exposure in the global Internet.

Our results motivate enhanced security with respect to international communications that may be of interest to foreign entities. We assume that strong encryption for such communications is already being implemented. Enhancements may be made in the area of hiding the communicating endpoints for communications in which this is relevant (e.g., communicating intelligence centers or military commands). The use of proxies and anonymous routing services can be assistive (although network throughput can then suffer and such services have had vulnerabilities). A remaining danger is that strongly encrypted traffic will be stored and then decrypted once quantum computers are accessible to nation states or until used cryptographic algorithms have been broken. Our findings on information exposure then promotes the changeover to quantum resistant encryption. 

Lastly, our work has shown that it is possible to model the communications between countries to determine the information exposure. Our work is thus a template for how others can perform this calculation for operational use. Using our approach, allied countries can evaluate their Internet communications and determine a lower bound on which other countries have the access to eavesdrop on their traffic. If that set of countries contains only allies, the risk of information exposure is diminished (but not eliminated). If not, additional security measures should be considered for highly sensitive data.


\section*{Acknowledgment} 
This work was partially accomplished under NIST Cooperative Agreement No.70NANB16H024 with the University of Maryland.
The authors would like to thank CAIDA and the University of Oregon for providing the data. 


\bibliographystyle{IEEEtran}
\bibliography{IEEEabrv,IEpaper}

\end{document}